\documentclass[noeprint,noshowpacs,nopreprintnumbers,twocolumn,prb,showpacs,superscriptaddress,amsmath,amssymb,citeautoscript,aps,10pt]{revtex4-1}
\usepackage{graphicx}
\usepackage{dcolumn}
\usepackage{color}
\usepackage{bm}
\usepackage[hidelinks]{hyperref}
\hypersetup{
    colorlinks,
    citecolor=blue,
    filecolor=blue,
    linkcolor=blue,
    urlcolor=blue
}

\newcommand{\ud}{\,\mathrm{d}}
\newcommand{\im}{\mathrm{i}}
\newcommand{\e}{\textrm{e}}

\DeclareMathOperator{\Tr}{Tr}

\DeclareMathOperator{\sgn}{sgn}

\DeclareMathOperator{\C}{\mathcal{C}}

\allowdisplaybreaks

\begin{document}

\title{Dynamical Phase Transitions in Topological Insulators}

\author{N.~Sedlmayr}
\email{sedlmayr@umcs.pl}
\affiliation{Department of Physics and Medical Engineering, Rzesz\'ow University of Technology, Poland}
\affiliation{Institute of Physics, M.~Curie-Sk{\l}odowska University, 20-031 Lublin, Poland}

\date{\today}

\begin{abstract}
The traditional concept of phase transitions has, in recent years, been widened in a number of interesting ways. The concept of a topological phase transition separating phases with a different ground state topology, rather than phases of different symmetries, has become a large widely studied field in its own right. Additionally an analogy between phase transitions, described by non-analyticities in the derivatives of the free energy, and non-analyticities which occur in dynamically evolving correlation functions has been drawn. These are called dynamical phase transitions and one is often now far from the equilibrium situation. In these short lecture notes we will give a brief overview of the history of these concepts, focusing in particular on the way in which dynamical phase transitions themselves can be used to shed light on topological phase transitions and topological phases. We will go on to focus, first, on the effect which the topologically protected edge states, which are one of the interesting consequences of topological phases, have on dynamical phase transitions. Second we will consider what happens in the experimentally relevant situations where the system begins either in a thermal state rather than the ground state, or exchanges particles with an external environment.
\end{abstract}

\maketitle


\section{Introduction}

Our current understanding of phase transitions owes much to the phenomenological theory of Landau.\cite{Landau1980,Sachdev2011} In this theory a second order, or continuous, phase transition is accompanied by a symmetry breaking across the transition, demonstrated by an order parameter. This is complemented by the earlier Ehrenfest classification, in which phase transitions are classified by non-analyticities which appear in derivatives of the free energy. Modern physics has added two new types of phase transition to these concepts: topological phase transitions and dynamical phase transitions.

A topological phase transition is accompanied by a change not in symmetry but rather in topology across the phase boundary.\cite{Hasan2010,Asboth2016} Band insulators, which possess a gap in their spectrum, can be classified by the topology of their band structure. If we consider a band insulator with a gap at zero energy, or a superconductor which has a quasi-particle gap at zero energy, then we can ascribe an integer $\mathbb{Z}$ or binary $\mathbb{Z}_2$ invariant to the negative energy bands. Note however that the total band structure will always have an invariant of zero. This invariant can only be changed by either closing the gap, or by changing the symmetry properties of the Hamiltonian\cite{Schnyder2009,Ryu2010,Chiu2016}. Typically this latter option is not considered and we are interested in the topological phase transitions in which the gap closes and opens, and the invariant changes, as a function of some parameter of the system with the symmetry properties of the Hamiltonian remaining the same. Such a phase transition can be contrasted with Landau's picture of symmetry breaking for continuous phase transitions. In this lecture we will focus purely on one dimensional (1D) topological insulators.

One of the most interesting and widely studied consequences of those phases in topological systems with non-zero topological invariants, which we will refer to generically as topologically non-trivial phase, is the existence of protected edge states which appear at the boundaries. There is a bulk-boundary correspondence\cite{Ryu2002} which proves that the bulk topological invariant determines the number of protected edge states which appear at the boundaries. These edge states have an exponentially small energy as a function of the system length, and are robust to disorder due to the bulk protection.

The second type of phase transition we are interested in is a dynamical phase transition. In this case an analogy is forged between the non-analytical behaviour of derivatives of the free energy, as in a continuous phase transition, and non-analytical behaviour in dynamical observables as a function of time. In particular we focus here on the quantum mechanical overlap between an initial state and a time evolved state as the observable, which is often called the Loschmidt amplitude.

The Loschmidt amplitude bears some relation to the fidelity, which is the overlap between two quantum states. Due to its universal scaling behaviour near quantum phase transitions this can be used to study phase transitions.\cite{Zanardi2006,Venuti2007,Rams2011,Sirker2010,Sirker2014a,Konig2016} For the 1D topological systems we are interested in it has been shown that the fidelity has universal finite size scaling behaviour\cite{Konig2016} and there are characteristic signatures in the fidelity which originate from the boundaries and demonstrate the existence or absence of the topologically protected edge states\cite{Sirker2014a}.

After some background on 1D topological insulators and superconductors, focusing on two particular examples, we will introduce dynamical phase transitions for these systems.\cite{Vajna2015,Konig2016,Sedlmayr2018,Zache2019} Following references \onlinecite{Vajna2015} and \onlinecite{Sedlmayr2018} we will then consider the particular properties of dynamical phase transitions when applied to topological systems, and the effect of the edge states. Finally we will look at several generalisations of the dynamical phase transitions to finite temperatures and open systems with particle loss or gain processes.

\subsection{Topological Phase Transitions}

We consider 1D Hamiltonians of the generic form
\begin{equation}
\label{1dham}
H=\sum_{ k}\Psi^\dagger_{ k}\mathcal{H}( k)\Psi_{ k}
\end{equation}
with
\begin{equation}
\label{1dhamp2}
\mathcal{H}(k)=\mathbf{d}_k \cdot {\bm\tau}\,,
\end{equation}
where ${\bm\tau}=(\bm\tau^x,\bm\tau^y,\bm\tau^z)$ are the Pauli matrices, which act in some subspace, $\mathbf{d}_k=(d^x_k,d^y_k,d^z_k)$, and $\Psi_{ k}$ are the appropriate creation or annihilation operators for that subspace. Below we will focus on two examples where this subspace will either be a physical lattice subspace when there are two particles in the unit cell, or particle-hole space for a superconductor described by a Bogoliubov-de Gennes Hamiltonian. Diagonalising $\mathbf{d}_k \cdot {\bm\tau}$ one finds $\mathbf{\tilde d}_k \cdot {\bm\tilde\tau}$ with $\mathbf{\tilde d}_k=(0,0,\epsilon_k)$. The pairs of eigenenergies $\pm \epsilon_k=\pm|\mathbf{d}_k|$ are a result of the particle-hole symmetry of the Hamiltonian.

In one dimension the topological invariant we are interested in is the winding number or Zak-Berry phase\cite{Berry1984,Zak1989,Ryu2006,Viyuela2014}. For a two band model such as \eqref{1dham} we can calculate the Zak-Berry phase for the lower energy band:
\begin{equation}\label{zak}
	\varphi=\im\int\ud k\langle u_k|\partial_ku_k\rangle\,,
\end{equation}
with the integral taken round the Brillouin zone and $|u_k\rangle$ being an eigenstate of the lower band: $H|u_k\rangle=-\epsilon_k|u_k\rangle$. This results in either $\mathbb{Z}$ or $\mathbb{Z}_2$ invariants, depending on the symmetries of the model\cite{Ryu2010}, and in turn this tells us how many topologically protected edge states will be present.\cite{Ryu2002,Teo2010} 

In 1D in the ``ten-fold way'' symmetry classification\cite{Ryu2010} we have three symmetry classes with ground states labeled by a $\mathbb{Z}$ topological invariant: AIII, BDI, and CII; and two labeled by a $\mathbb{Z}_2$ topological invariant: D, and DIII. We will focus on examples in the BDI class, which have particle-hole symmetry, a form of time reversal symmetry and chiral symmetry, which is in fact the combination of the previous two. Hamiltonians with particle-hole symmetry obey the anti-commutation relation $\{\C,H\}=0$ with $\C$ the unitary particle-hole operator satisfying $\C^2=1$. Secondly the time reversal operator is $\mathcal{T}$, with $\mathcal{T}^2=1$, and we have $[\mathcal{T},H]=0$. We demand that $\{\C,\mathcal{T}\}=0$ and note that the chiral symmetry is simply $\{\mathcal{T}\C,H\}=0$.

Due to the particle-hole symmetry it is always possible to make a \emph{momentum independent} rotation to a basis in which, for example, $d^z_k=0$ in equation \eqref{1dhamp2}. The eigenstates for the negative energy band are then
\begin{equation}
	|u_k\rangle=\frac{1}{\sqrt{2}}\begin{pmatrix}
		1\\-\e^{-\im\phi_k}\,,
	\end{pmatrix}
\end{equation}
with
\begin{equation}
	\e^{-\im\phi_k}=\frac{d^x_k-\im d^y_k}{\sqrt{(d^x_k)^2+(d^y_k)^2}}\,.
\end{equation}
The Zak-Berry phase then becomes
\begin{equation}\label{zak2}
	\varphi=\frac{1}{2}\int\ud k\partial_k\phi_k=\pi \nu\,,\textrm{ with }\nu\in\mathbb{Z}\,.
\end{equation}
This last part follows from considering how many times the line of $\e^{\im\phi_k}$ encloses zero in the complex plane for $k:0\to 2\pi$. This number cannot change unless $d^x_k=d^y_k=0$, which is the gap closing condition.

\begin{figure}
	\begin{center}
		\includegraphics[width=\columnwidth]{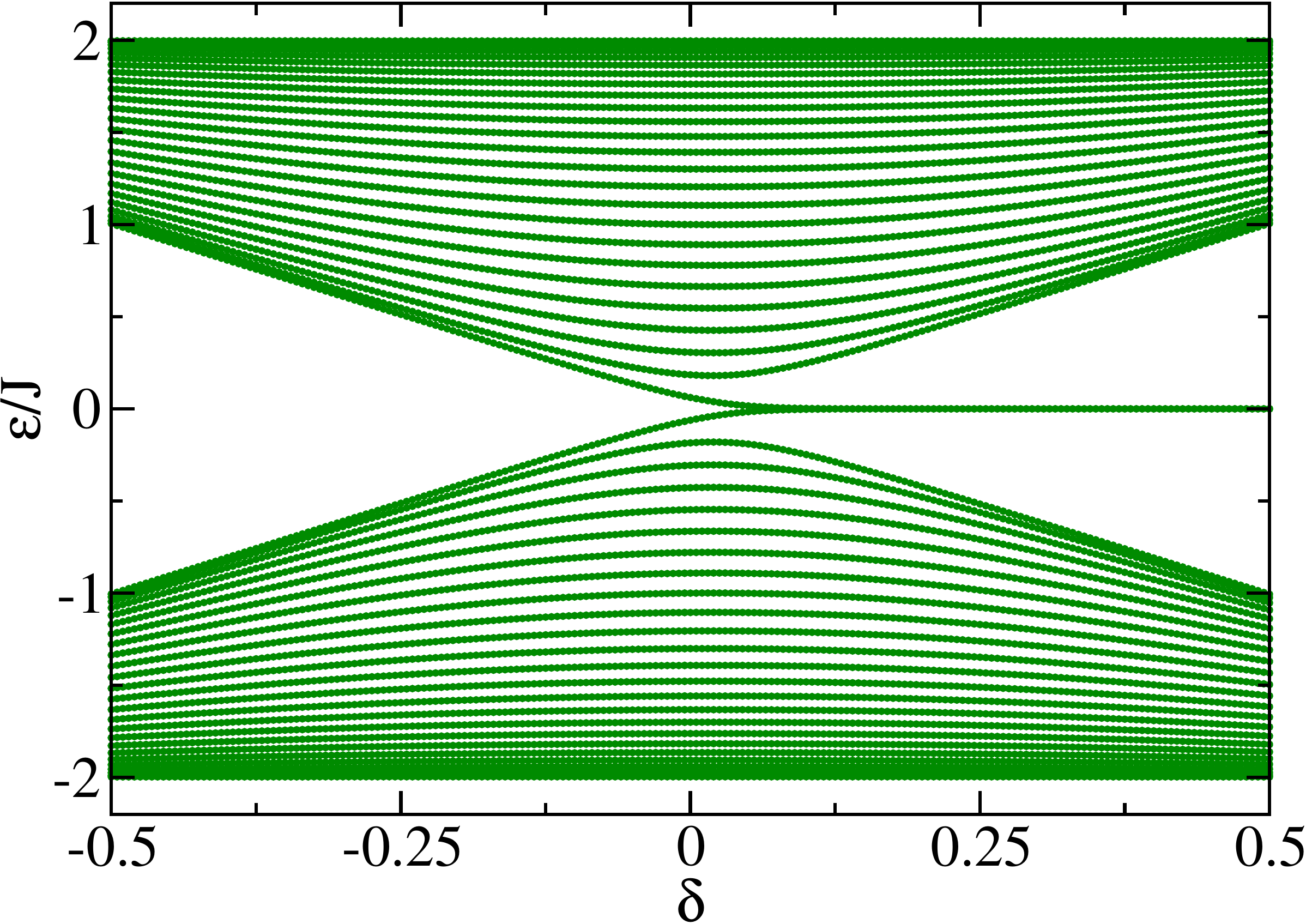}
		\caption{An example of a topological phase transition in the SSH model, see equation \eqref{ssh_model}, showing the spectrum as a function of the dimerisation $\delta$. The bulk gap closes and opens at $\delta=0$ and for $\delta>0$ the systems is in a topologically non-trivial phase as demonstrated by the existence of the zero energy edge states. Calculated for a system of size $N=50$ with OBCs.}
		\label{fig_spectrum}
	\end{center}
\end{figure}

We consider two exemplary one dimensional (1D) topological insulators/superconductors, the Su-Schrieffer-Heeger\cite{Su1980} (SSH) model and the long range Kitaev chain\cite{Kitaev2001,DeGottardi2013}. Both of these models are in the BDI symmetry class with particle-hole, time reversal, and chiral symmetries. We will analyse these models both in the case where they have periodic boundary conditions (PBCs), to investigate the bulk properties, and for open boundary conditions (OBCs), to consider the role of the boundaries and the topologically protected edge modes.

The SSH model is a simple dimerised chain, originally introduced to describe polymers like polyacetyline\cite{Su1980} and has the Hamiltonian
\begin{equation}
\label{ssh_model}
H=-J\sum_{j}\left[(1+\delta\e^{i\pi j})c^\dagger_jc_{j+1}+\textrm{H.c.}\right]\,.
\end{equation}
$c^\dagger_j$ creates a fermionic particle at site $j$. $J$ is the average hopping integral and $\delta$ the strength of the dimerisation. At $\delta=0$ this system becomes critical and the gap closes. This critical point separates the topologically trivial, $\delta<0$, and non-trivial, $\delta>0$, phases. For an example see figure \ref{fig_spectrum}. A straightforward Fourier transform and subspace rotation will transform \eqref{ssh_model} into \eqref{1dham} with
\begin{equation}
\label{d_SSH}
\mathbf{d}_k=\begin{pmatrix}
-2J\cos k,2J\delta\sin k,0
\end{pmatrix}\,.
\end{equation}
Due to the definition of the unit cell implicit here in this case the Brillouin zone is actually defined as $k:0\to\pi$ and the Zak-Berry phase becomes
\begin{equation}
	\varphi=\frac{\pi}{2}\sgn\delta\,,
\end{equation}
with $\delta>0$ being the non-trivial phase.

In order to consider more general scenarios we also consider a Kitaev chain of $M$ sites with long-range hopping terms: 
\begin{eqnarray}\label{khlr}
H&=&\sum_{i,j}\Psi^\dagger_{i}\left(
\Delta_{|i-j|}\im{\bm\tau}^y-J_{|i-j|}{\bm\tau}^z\right)\Psi_{j+1}+\textrm{H.c.}\nonumber\\&&-\mu\sum_{j}\Psi^\dagger_{j}{\bm\tau}^z\Psi_{j}\,.
\end{eqnarray}
This is a Bogoliobov-de Gennes Hamiltonian for spinless particles with ${\bm\tau}$ representing the particle-hole space. $\Psi^\dagger_{j}=(c^\dagger_{j},c_{j})$ and $c_{j}^{(\dagger)}$ annihilates (creates) a fermionic particle on site $j$. $\Delta_{|i-j|}$ and $J_{|i-j|}$ are a p-wave like superconducting pairing and hopping term respectively. A Fourier transform and rotation will give us \eqref{1dham} with
$\Psi^\dagger_k=(c^\dagger_k,c_{-k})$ and
\begin{equation}
\label{d_Kit}
\mathbf{d}_k=\sum_{m} \begin{pmatrix}
-2J_m\cos[mk],2\Delta_m\sin[mk],0
\end{pmatrix}
-\begin{pmatrix}
\mu,0,0
\end{pmatrix}
\,.
\end{equation}
In this case one can see that the Zak-Berry phase can in principle now result in invariants of any integer which allows us to consider cases with many protected edge states. Here we will limit the long range terms to $\Delta_{m\geq4}=J_{m\geq4}=0$.

\subsection{Dynamical Phase Transitions}

The concept of dynamical phase transitions introduced by Heyl, Polkovnikov, and Kehrein in 2013\cite{Heyl2013} is based upon an analogy between the equilibrium partition function, namely
\begin{equation}
	Z(\beta)=\Tr\e^{-\beta H}
\end{equation}
for a Hamiltonian $H$ at inverse temperature $\beta$, and the overlap between an initial state $|\psi_0\rangle$ and its time evolved counterpart $\e^{\im Ht}|\psi_0\rangle$:
\begin{equation}
	L(t)=\left\langle\psi_0\left|\e^{-\im Ht}\right|\psi_0\right\rangle\,.
\end{equation}
This latter quantity is the Loschmidt amplitude. Much like the non-analyticities as a function of $\beta$ which accompany an equilibrium phase transition, the Loschmidt amplitude can become non-analytic as a function of complex time $t$. At these times, referred to as Fisher zeroes, $L(t)$ vanishes. When the Fisher zeros cross the real time axis then a dynamical phase transition occurs.

The free energy for the partition function also has a counterpart for the Loschmidt echo and we can introduce the return rate
\begin{equation}\label{returnrate}
	l(t)=-\frac{1}{N}\ln |L(t)|\,,
\end{equation}
where $N$ is the system size. We note that this avoids issues associated with the expected Anderson orthogonality catastrophe for a many body system, in which the overlap between the initial state and the time evolved state will become exponentially small in the thermodynamic limit.

As is usual for dynamical phase transitions we focus on a particular form of non-equilibrium dynamics known as a quench. In a quench the system is first prepared in the many-body ground state $|\psi_0\rangle$ of a Hamiltonian $H_0$. This is then time evolved by a different Hamiltonian $H_1$. Typically $H_0$ and $H_1$ differ by a global parameter. For example we will consider quenches in which $H_0$ is the SSH model with $\delta<0$ and $H_1$ is the SSH model with $\delta>0$, a quench across the topological phase transition.

\begin{figure}
	\begin{center}
		\includegraphics[width=\columnwidth]{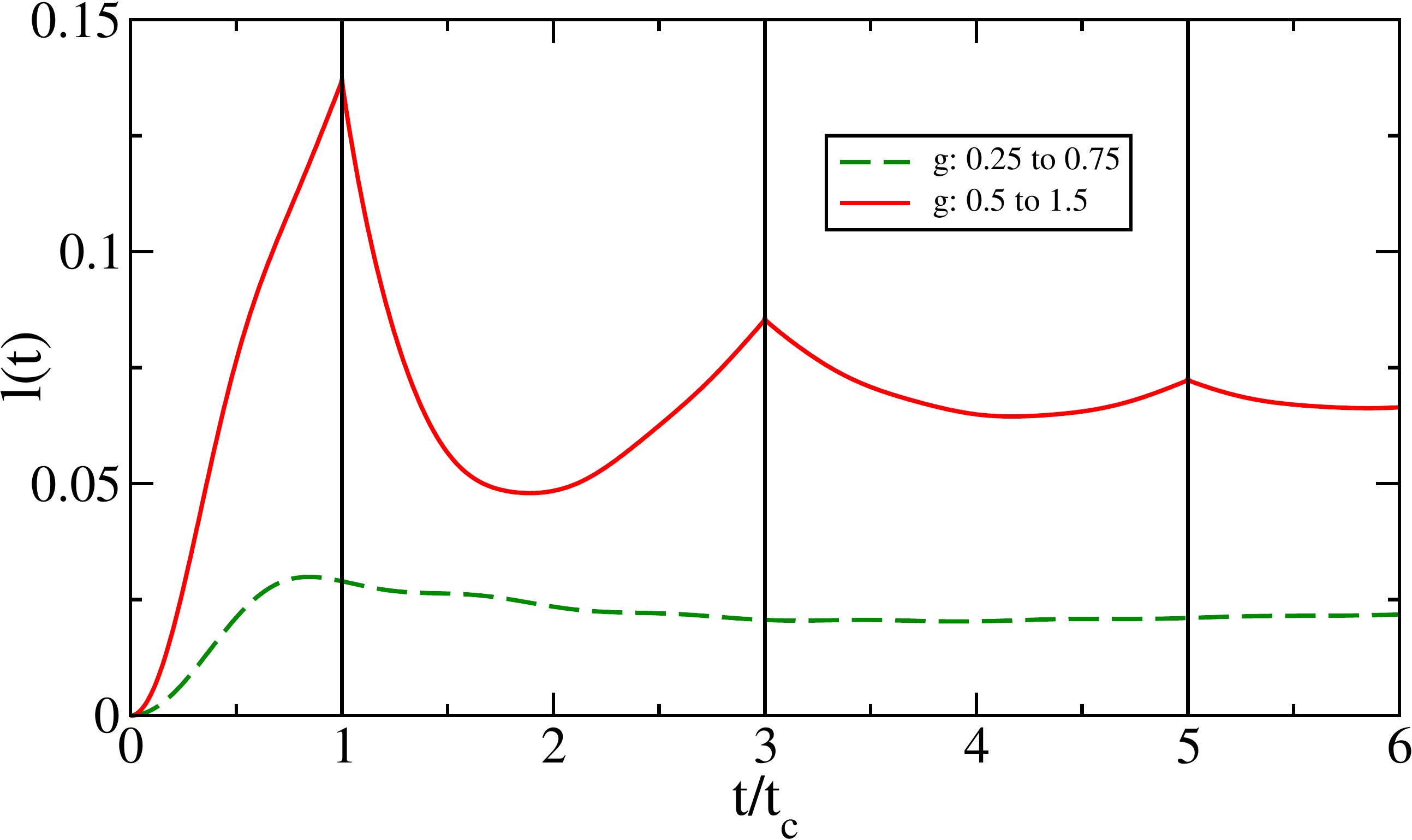}
		\caption{A dynamical phase transition in the Ising model, equation \eqref{Ising}, for a quench across the equilibrium phase transition ($g:0.5\to1.5$) and inside a phase ($g:0.25\to0.75$). Only the quench across the phase transition shows the cusps in the return rate associated with a dynamical phase transition. Both cases are scaled by the critical time for the quench $g:0.5\to1.5$.}
		\label{fig_ising_dpt}
	\end{center}
\end{figure}

It was first shown that the Ising model undergoes dynamical phase transitions when quenching across its equilibrium phase boundary\cite{Heyl2013} and soon generalised to more models.\cite{Karrasch2013} The Ising model is a 1D spin-$\frac{1}{2}$ chain with Hamiltonian
\begin{equation}\label{Ising}
	H(g)=-\frac{1}{2}\sum_{i}{\bm\sigma}^z_i{\bm\sigma}^z_{i+1}+\frac{g}{2}\sum_{i=1}^N{\bm\sigma}^x_i\,.
\end{equation}
This has a phase transition for an applied magnetic field $g=1$. Note that this model can be mapped to a special case of the Kitaev chain, equation \eqref{khlr}, with nearest-neighbour hopping only and $\mu=-g/2$ and $J_1=1/4=-\Delta_1$. At critical times $t_n=(2n+1)t_c$, where $n=0,1,2,\ldots$, the Loschmidt amplitude becomes zero $L(t_n)=0$ and hence the return rate diverges. The critical time $t_c$ can be calculated, and for a quench from an initial state which is the ground sate of $H(g_0)$, and time evolution with $H(g_1)$ one finds
\begin{equation}
	t_c=\pi\sqrt{\frac{g_1+g_0}{(g_1-g_0)(g_1^2-1)}}\,.
\end{equation}
Two examples are shown in figure \ref{fig_ising_dpt}, when the system is quenched across the equilibrium phase boundary at $g=1$ then there are dynamical phase transitions. However when it is quenched within an equilibrium phase then there are no dynamical phase transitions.

This lead to the belief that perhaps there was a one-to-one relation between the dynamical phase transitions and whether the quenches crossed the equilibrium phase boundary. However further work on more complicated systems demonstrated that no such direct relation exists.\cite{Vajna2014} For the XY chain in a transverse magnetic field it was demonstrated that dynamical phase transitions can occur with and without crossing equilibrium phase boundaries, and that crossing an equilibrium phase boundary with the quench does not necessarily imply a dynamical phase boundary.\cite{Vajna2014} Due to the potential richness of the non-equilibrium physics involved this is perhaps not very surprising. However for the case we consider here of 1D topological insulators there are some simple statements that can be made.

\section{Dynamical Phase Transitions in 1D Topological Insulators}

The Loschmidt amplitude for quenches in topological insulators with Hamiltonians as given by \ref{1dham}, can be easily calculated for periodic boundary conditions. One finds\cite{Vajna2015}
\begin{equation}\label{tilos}
	L(t)=\prod_k\left[\cos(\epsilon^1_kt)+\im\hat{\mathbf{d}}^0_k\cdot\hat{\mathbf{d}}^1_k\sin(\epsilon^1_kt)\right]\,,
\end{equation}
where $\mathbf{d}^{0,1}_k$ describes the initial ground state or the time evolving Hamiltonian respectively. Furthermore we define $\hat{\mathbf{d}}^{0,1}_k=\mathbf{d}^{0,1}_k/|\mathbf{d}^{0,1}_k|$ and $\epsilon^1_k$ as the positive eigenenergy of the time-evolving Hamiltonian. The product runs over the whole of the filled negative energy band.

In this case one can see that the critical times occur if there is a critical momentum satisfying
\begin{equation}
\label{tn3}
\hat{\mathbf{d}}^0_{k^*}\cdot\hat{\mathbf{d}}^1_{k^*}=0 \, .
\end{equation}
In that case
\begin{equation}
\label{tn2}
t_n=\frac{\pi}{2\epsilon^1_{k^*}}\left(2n+1\right)\,, \textrm{ where  }n=0,1,2,\ldots\,.
\end{equation}
Furthermore one can make the statement that if $\mathbf{d}^0_{k}$ and $\mathbf{d}^1_{k}$ belong to different topological phases there must be a solution to equation \eqref{tn3}.\cite{Vajna2015} For the long range Kitaev model there can be multiple solutions to equation \eqref{tn3}, and the number of critical momenta appears to be related to the change in the topological invariant between the ground state and the time evolving Hamiltonian.\cite{Sedlmayr2018a}

\begin{figure}
	\begin{center}
		\includegraphics[width=\columnwidth,height=5.05cm]{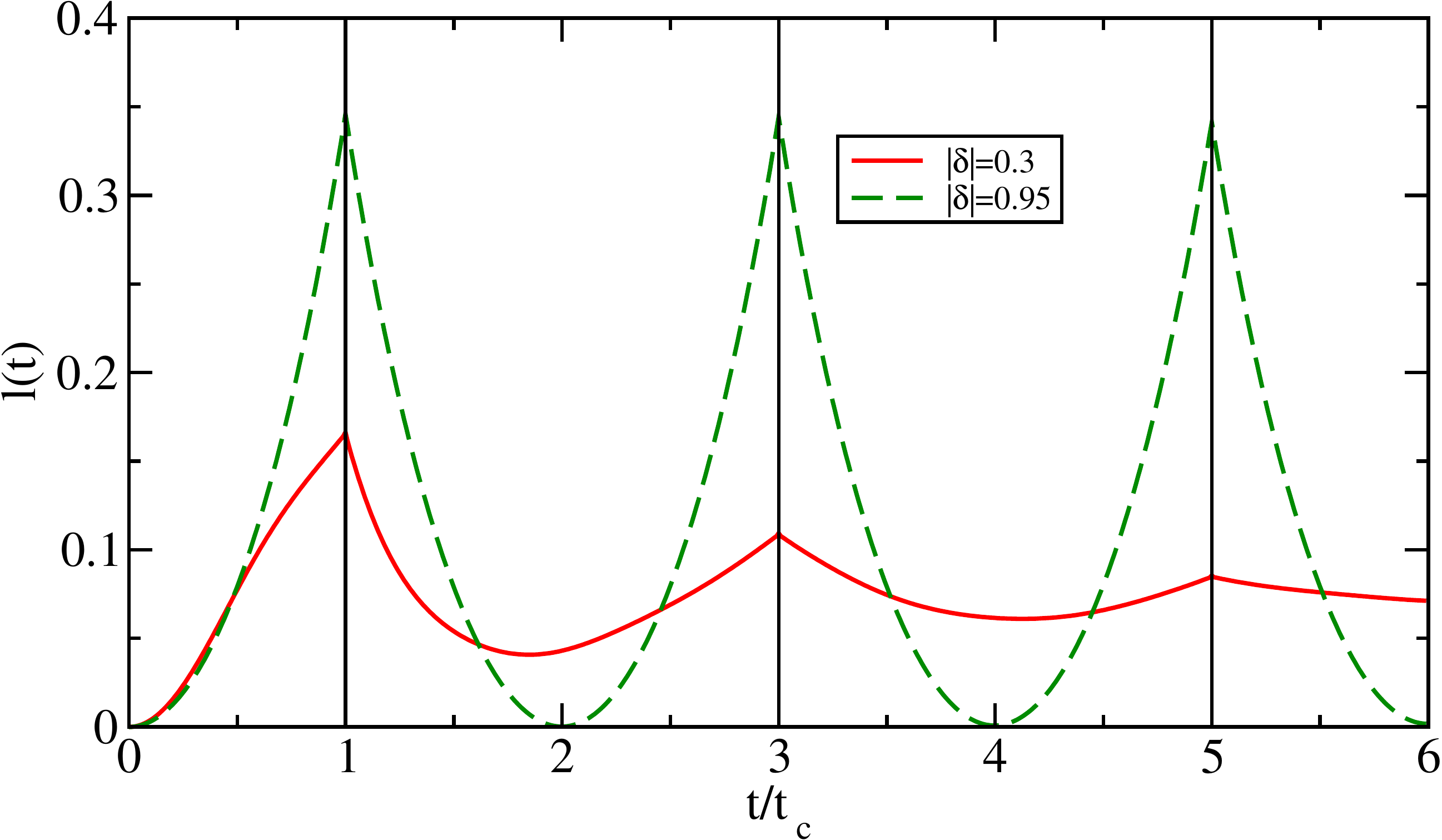}
		\caption{A dynamical phase transition in the SSH model, equation \eqref{ssh_model}, for quench across the topological phase transition: $\delta:|-\delta'|\to|\delta'|$ with $\delta'=0.3,-0.95$. Due to the symmetry of this model in the bulk there is no difference between these quenches and for $\delta:|\delta'|\to-|\delta'|$.}
		\label{fig_ssh_dpt}
	\end{center}
\end{figure}

\begin{figure}
	\begin{center}
		\includegraphics[width=\columnwidth,height=5.05cm]{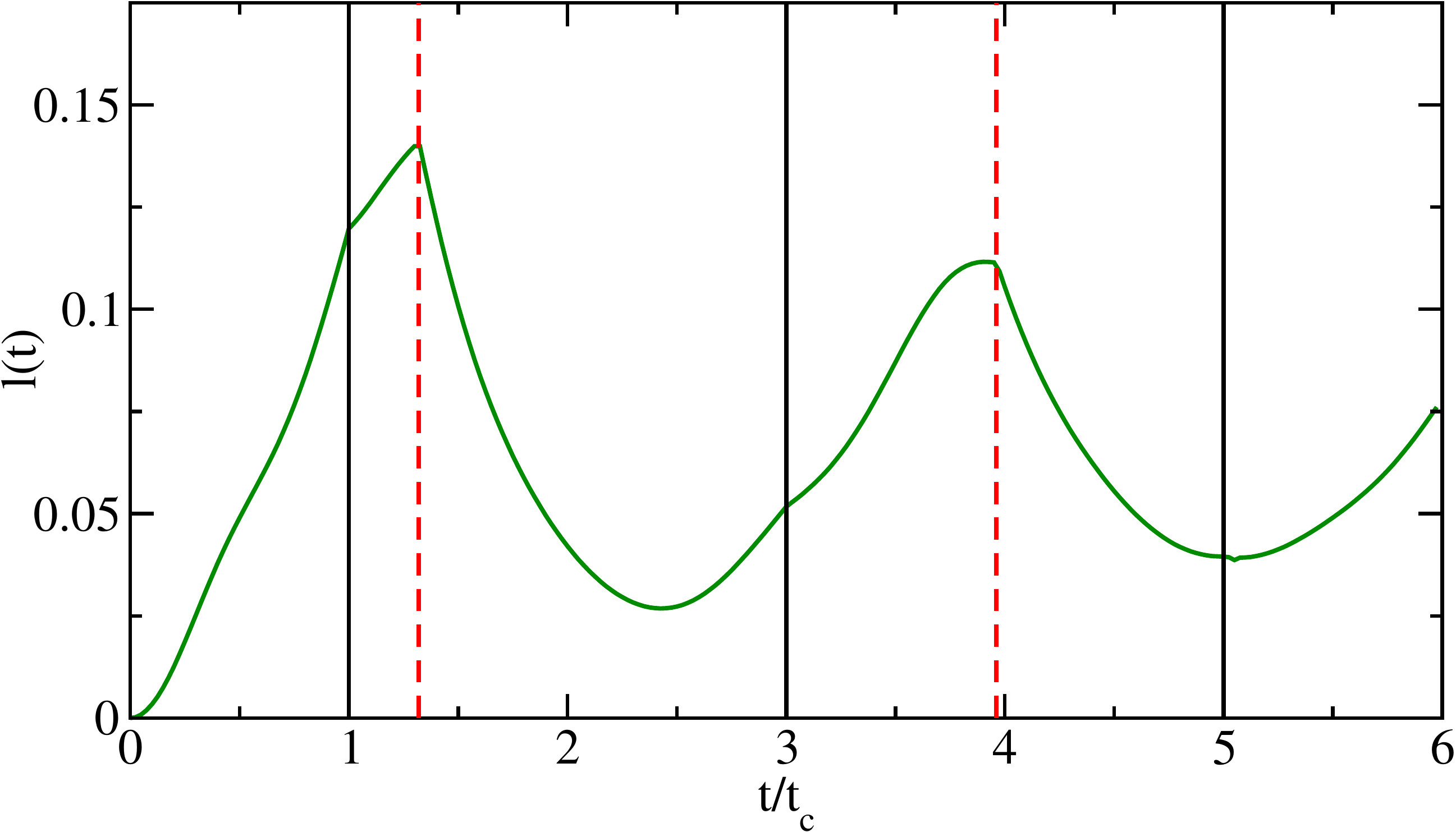}
		\caption{A dynamical phase transition in the long range Kitaev model, equation \eqref{khlr}, for quench across a topological phase transition with topological invariants 1 to 3. In this case there are two critical times at which dynamical phase transitions occur. The vertical black and dashed red lines how the critical times.}
		\label{fig_kit_dpt}
	\end{center}
\end{figure}

As an example in figure \ref{fig_ssh_dpt} we show two quenches across the topological phase transition for the SSH model equation \eqref{ssh_model}. In figure \ref{fig_kit_dpt} a dynamical phase transition with two critical times is shown. Here we have used equation \eqref{khlr} with the quench from $\nu=1$ with $\vec J=(1,-2,2)$, $\mu=2$, and $\vec\Delta=(1.3,-0.6,0.6)$ to $\nu=3$ with $\vec J=(1,-2,2)$, $\mu=0.1$, and $\vec\Delta=(0.45,-0.9,1.35)$.\cite{Sedlmayr2018a}

\subsection{Boundary Contributions}

One of the interesting consequences of the bulk topology of a topological insulator is the existence of the protected edge states. Here we will review what effect they have on the return rate. Once we consider open systems with edges where these states can exist we no longer have momentum as a good quantum number and hence we can not use equation \eqref{tilos}. Instead one can use the following formalism:\cite{Levitov1996,Klich2003,Rossini2007,Sedlmayr2018a}
\begin{equation}
\label{rle}
L(t)=\det\mathbf{M}\equiv\det\left[1-\mathbf{\C}+\mathbf{\C}\e^{\im {\bm H}_1t}\right]\,.
\end{equation}
where the correlation matrix $\mathbf{\C}$ for the initial state is $\C_{ij}=\langle\psi_0|\Psi^\dagger_i\Psi_j|\psi_0\rangle$.

\begin{figure}
	\begin{center}
		\includegraphics[width=\columnwidth]{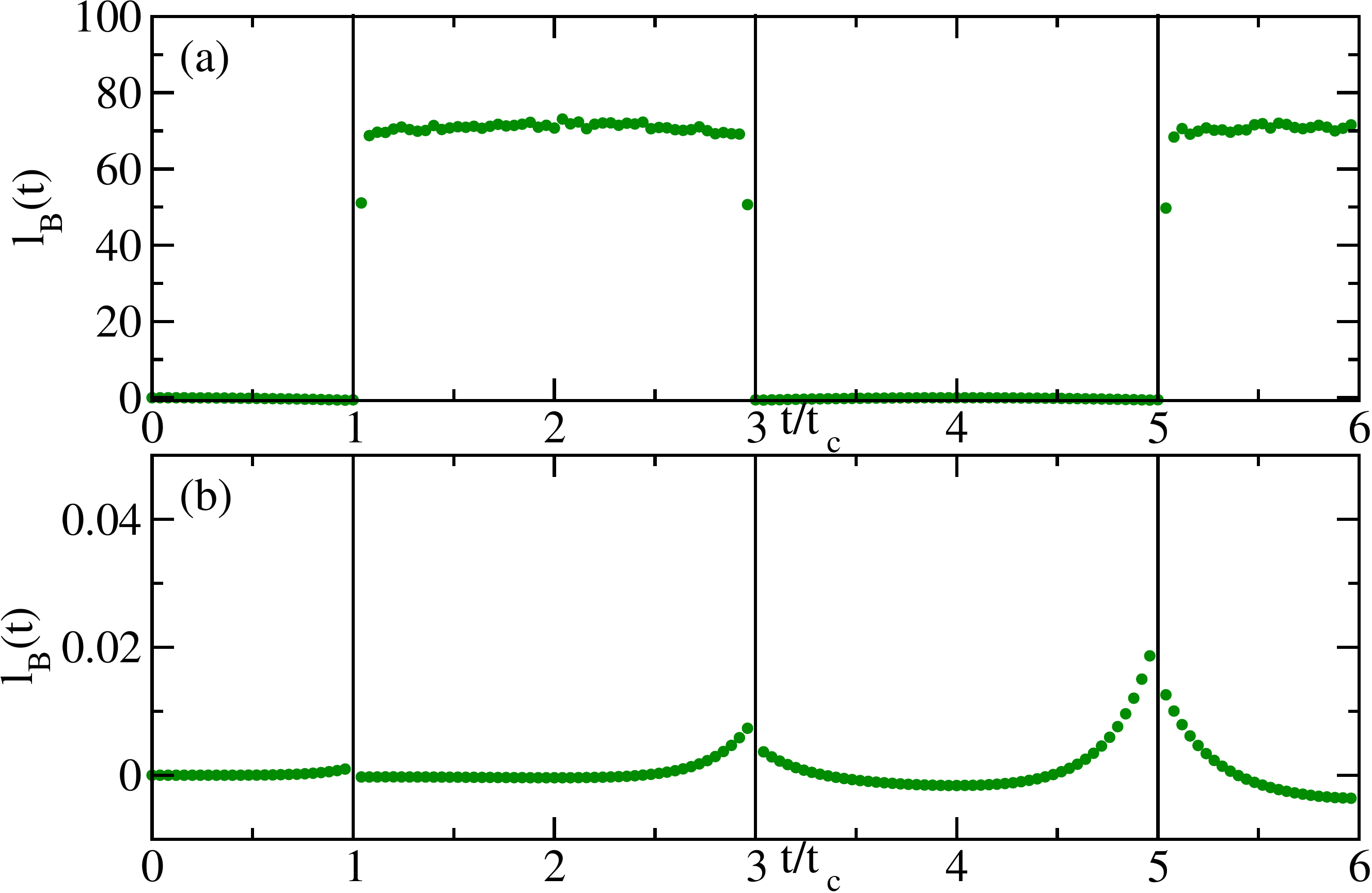}
		\caption{The boundary contribution to the return rate for quenches in the SSH model with (a) $\delta:-0.95\to0.95$, i.e.~from a topologically trivial to a topologically non-trivial phase, and (b) $\delta:0.95\to-0.95$, i.e.~from a topologically non-trivial to a topologically trivial phase\cite{Sedlmayr2018a}. }
		\label{fig_ssh_b_dpt}
	\end{center}
\end{figure}

To extract the effects of the edge states on the return rate one must perform a finite size scaling analysis for the boundary term $l_B(t)$:
\begin{equation}
	l_N(t)\sim l(t)+\frac{1}{N}l_B(t)\,.
\end{equation}
The boundary return rate demonstrates very different behaviour depending on the direction of the quench. We will focus on the case where $|\delta|=0.95$. In this strong dimerisation limit the effects are most clear. Figure \ref{fig_ssh_b_dpt} shows the strong asymmetry in the boundary term, which can be extracted from finite size scaling, for quenches in the two directions across the topological phase boundary.\cite{Sedlmayr2018a} The very large jumps in \ref{fig_ssh_b_dpt}(a) are due to the role of the topologically protected edge states during time evolution.

\begin{figure}
	\begin{center}
		\includegraphics[width=\columnwidth]{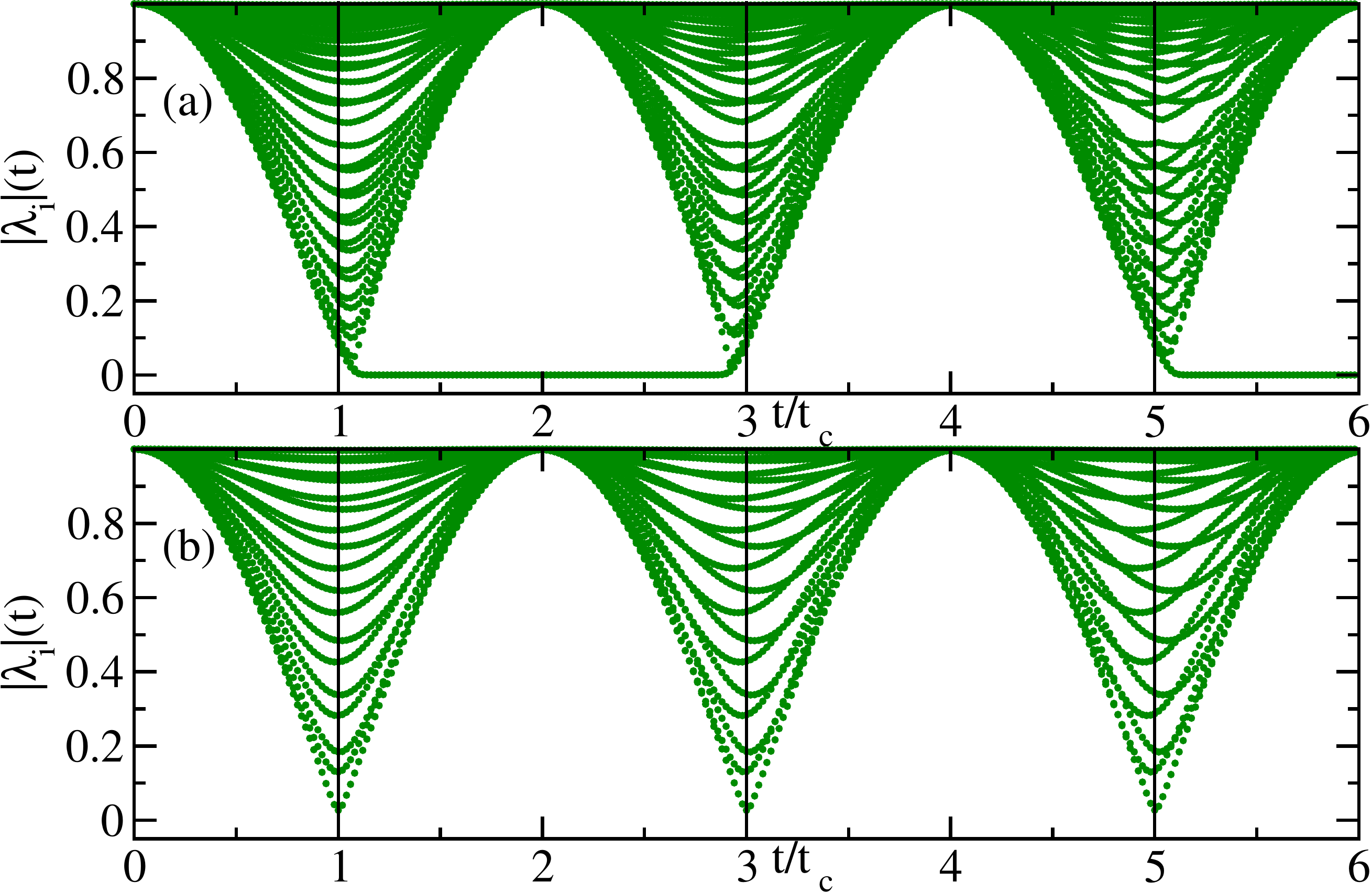}
		\caption{The eigenvalues of the Loschmidt matrix $\lambda_i(t)$ for (a) $\delta:-0.95\to0.95$, i.e.~from a topologically trivial to a topologically non-trivial phase, and (b) $\delta:0.95\to-0.95$, i.e.~from a topologically non-trivial to a topologically trivial phase.\cite{Sedlmayr2018a} In case (a) pairs of eigenvalues become pinned to zero between alternating critical times, causing the jumps which can be seen in the boundary return rate $l_B(t)$ in figure \ref{fig_ssh_b_dpt}(a). The system size is $N=40$.}
		\label{fig_ssh_eigenvalues}
	\end{center}
\end{figure}

The origin of the large jumps in the boundary return rate are caused by eigenvalues of the Loschmidt matrix, $\lambda_i(t)$, which become pinned to zero between critical times, see figure \ref{fig_ssh_eigenvalues}. In fact a direct comparison between the contribution of these two eigenvalues to the return rate and the boundary return rate shows remarkably good agreement.\cite{Sedlmayr2018}

\subsection{A Potential Relation to Entanglement Entropy}

\begin{figure}
	\begin{center}
		\includegraphics[width=\columnwidth]{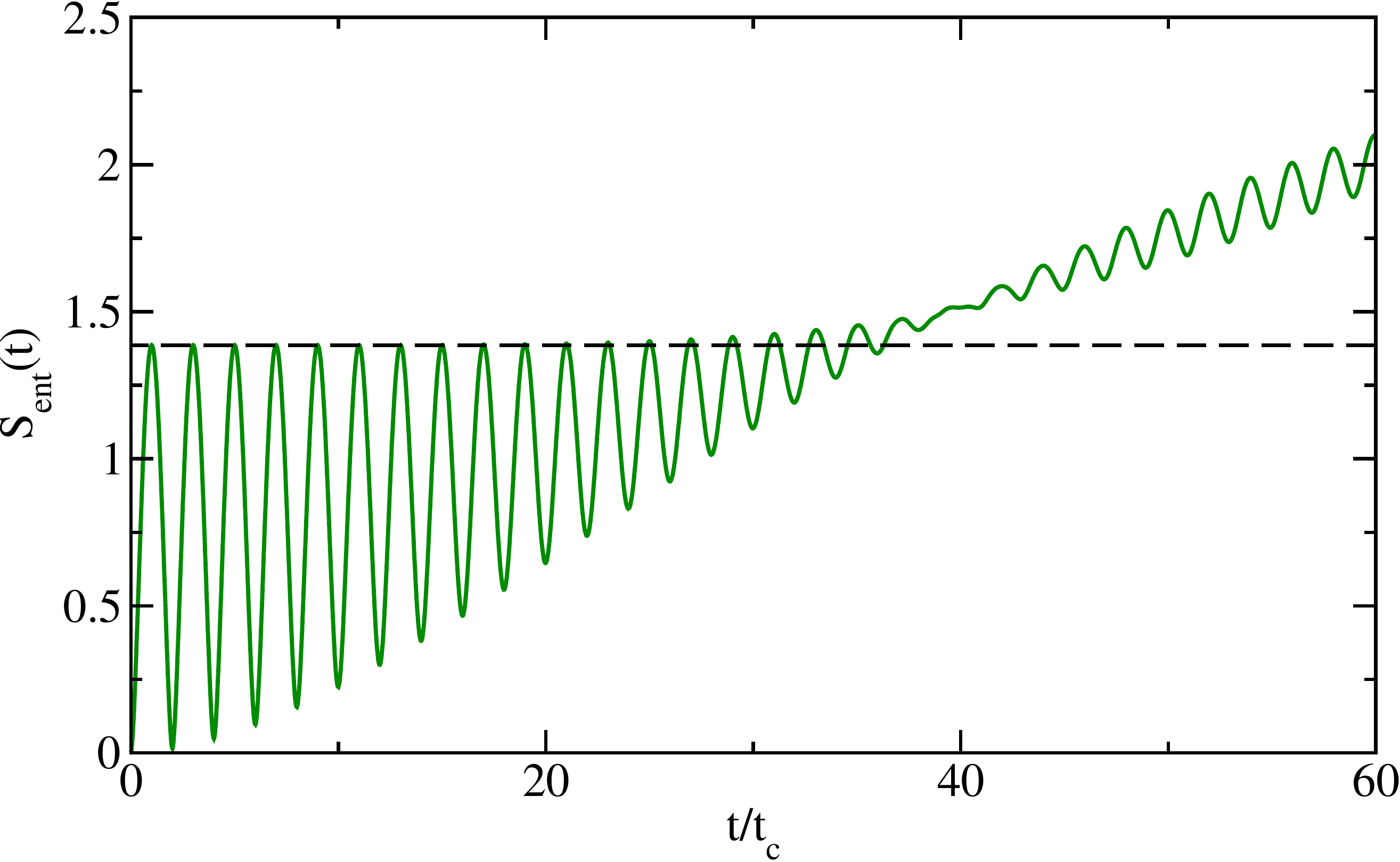}
		\caption{The entanglement entropy for a quench in the SSH model as in figure \ref{fig_ssh_b_dpt}(a). The solid dashed line is the entanglement entropy for cutting two dimers. At short times the system oscillates between this situation and $A$ and $B$ being unentangled as entanglement slowly builds up.\cite{Sedlmayr2018a} The system size is $N=32$.}
		\label{fig_ssh_ent}
	\end{center}
\end{figure}

Curiously it appears that the critical time of the dynamical phase transition plays an important role also in the time evolution of entanglement entropy which follows a quench.\cite{Sedlmayr2018} Entanglement entropy is the von-Neumann entropy of a reduced density matrix $\rho_A(t)=\Tr_B |\Psi(t)\rangle\langle\Psi(t)|$ defined as
\begin{equation}
 S_{\textrm{ent}}(t) =-\Tr \{\rho_A(t) \ln \rho_A(t)\}\,,
\end{equation}
with $|\Psi(t)\rangle = \text{e}^{-iH_1t}|\psi_0\rangle$ being the
time-evolved state. The system has been divided up into two blocks equally sized blocks $A$ and $B$. Figure \ref{fig_ssh_ent} shows that the entanglement entropy oscillates at the exact frequency of the critical time of the dynamical phase transition. The reason for this remains currently unclear, but suggests some deeper connection between these phenomena.

\section{Finite Temperatures}

The preceding sections focused exclusively on the case where the system is initially in a ground state. In any real experimental situation\cite{Jurcevic2017,Flaschner2018} the system is likely to be at a finite, if small, temperature. It is therefore interesting to generalise the concept of dynamical phase transitions to finite temperatures and density matrices.\cite{Abeling2016,Heyl2017,Sedlmayr2018a,Mera2017,Lang2018,Lang2018a} There is no unique way in which one may want to make such a generalisation and several versions have been considered in the literature. Here we will review several of these generalisations for the Loschmidt echo, which is the absolute value of the Loschmidt amplitude. For any generalisation we can define a return rate as in equation \eqref{returnrate}.

The first example we consider is of the Loschmidt echo as a metric in Hilbert space.\cite{Sedlmayr2018a} To define a metric for density matrices $\rho(t)$ the Loschmidt amplitude ${\cal L}_\rho(\rho(0),\rho(t))$ needs to satisfy
\begin{itemize}
\item[i.] $0\leq |{\cal L}_\rho(\rho(0),\rho(t))|\leq1$ and $|{\cal L}_\rho(\rho(0),\rho(0))|=1$,
\item[ii.] $|{\cal L}_\rho(\rho(0),\rho(t))|=1$ iff $\rho(0)=\rho(t)$, and
\item[iii.] $|{\cal L}_\rho(\rho(0),\rho(t))|=|{\cal L}_\rho(\rho(t),\rho(0))|$.
\end{itemize}
Following the definition of fidelity for density matrices\cite{Bures1969,Uhlmann1976,Jozsa1994} leads to\cite{Zanardi2007a,CamposVenuti2011}
\begin{equation}
\label{LoT}
{\cal L}_\rho(t)\equiv |{\cal L}_\rho(\rho(0),\rho(t))|=\Tr\sqrt{\sqrt{\rho(0)}\rho(t)\sqrt{\rho(0)}}\,.
\end{equation}
Despite this looking like a rather unwieldy expression for Hamiltonians of the form \eqref{1dham}, and with $\rho(0)$ being the canonical density matrix at a finite temperature $T=\beta^{-1}$, ${\cal L}_\rho(t)$ can be calculated exactly analytically\cite{Sedlmayr2018a} or in a semiclassical approximation.\cite{Lang2018}

An alternative inspired by the experiments\cite{Jurcevic2017} which correspond to an initial density matrix is time evolved and then projected onto a pure state\cite{Sedlmayr2018a} is
\begin{equation}
\label{Lproj}
|{\cal L}_p(t)|^2 = \frac{\langle
\Psi_0^0|\rho(t)|\Psi_0^0\rangle}{\langle\Psi_0^0|\rho(0)|\Psi_0^0\rangle}
= \sum_n \frac{p_n}{p_0} |\langle \Psi_0^0
|\e^{-iHt}|\Psi_n^0\rangle|^2 \, .
\end{equation}
Alternatively one could consider averaging over the pure state Loschmidt amplitudes with a weighting determined by an initial density matrix:\cite{Heyl2017}
\begin{equation}
\label{Lav}
{\cal L}_{\textrm{av}}=\Tr\left\{\rho(0)S(t)\right\}=\sum_n p_n \langle \Psi^0_n|\e^{-iH_1 t}|\Psi^0_n\rangle\,,
\end{equation}
with $S(t)$ being the time-evolution operator.

Another experimentally motivated expression is to relate the Loschmidt echo to the characteristic function of work.\cite{Talkner2007,Abeling2016} In that case the Loschmidt amplitude is given by
\begin{eqnarray}
\label{tildeLav}
\tilde {\cal L}_{\textrm{av}} &=& \frac{1}{Z}\Tr\left\{\e^{iH_1t}\e^{-iH_0t}\e^{-\beta H_0}\right\} \\
&=& \frac{1}{Z}\sum_n \e^{-(\beta+it)E_n^0} \langle \Psi^0_n|\e^{iH_1t}|\Psi^0_n\rangle \nonumber \, ,
\end{eqnarray}
and describes a thermal average over the Loschmidt echo of pure states.

\begin{figure}
	\begin{center}
		\includegraphics[width=\columnwidth]{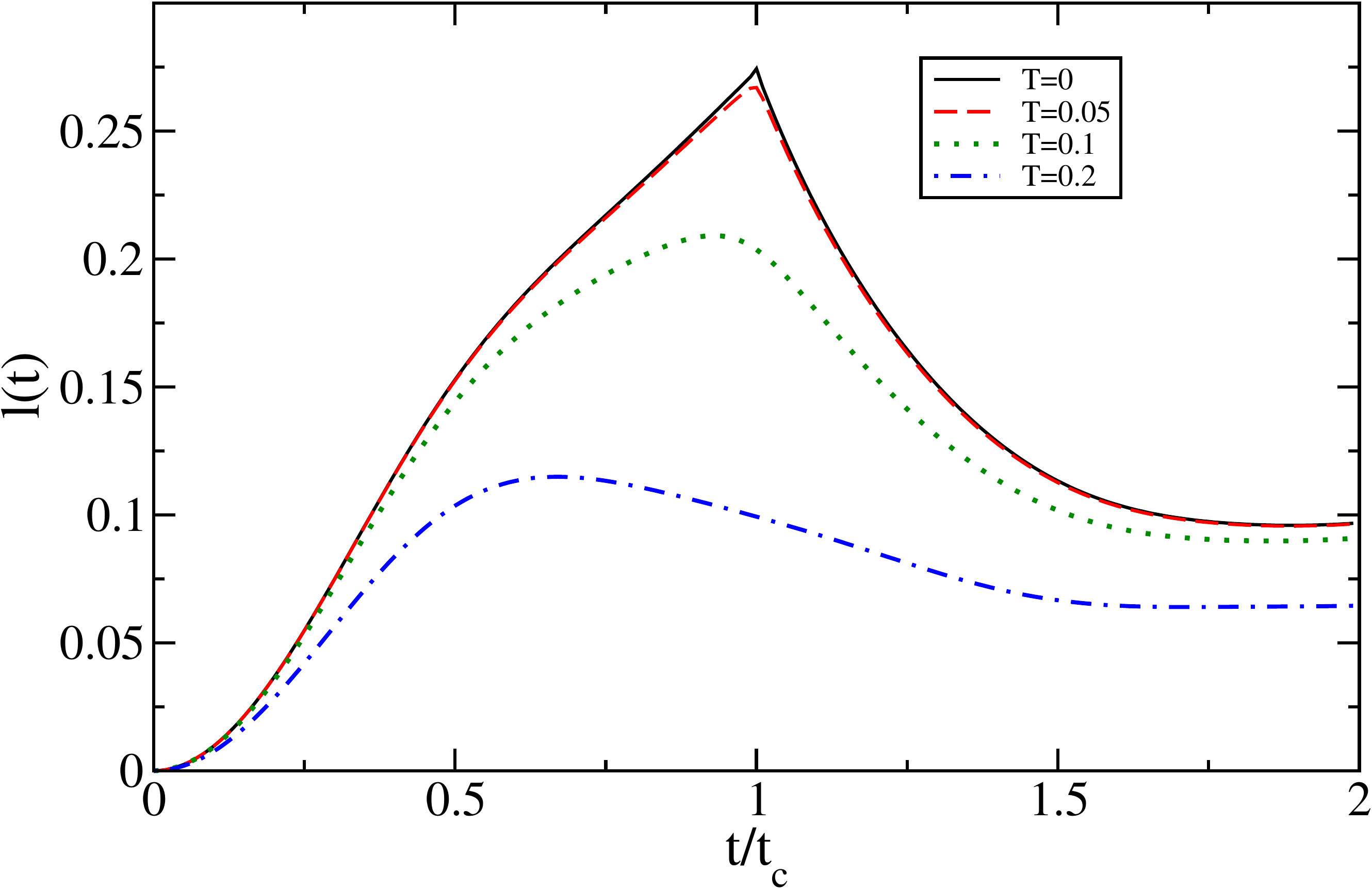}
		\caption{An example of the smoothing off of the cusp in the return rate at finite temperatures. For a quench in the Ising chain from $g=0.5$ to $g=1.5$ as in figure \ref{fig_ising_dpt}. The critical time is calculated for the zero temperature case. Here we have used the metric generalisation for the Loschmidt echo $|{\cal L}_\rho(t)|$, see equation \eqref{LoT}~\cite{Sedlmayr2018}. }
		\label{fig_finite_t}
	\end{center}
\end{figure}

Generically it is found for these generalisations that finite temperatures have the effect of destroying the cusp in the return rate, see figure \ref{fig_finite_t} for an example. One exception is for ${\cal L}_{\textrm{av}}$ in the effectively finely tuned case where the occupation of every momentum mode is conserved by a generalised Gibbs ensemble.\cite{Heyl2017}

It is also possible to include the loss and creation of particles during the time evolution by solving the Lindblad equation for the time evolution of the density matrix in the Born-Markov approximation. As for most finite temperature cases one also seen finds that the cusps which signature the dynamical phase transitions are removed, except for very finely tuned cases.\cite{Sedlmayr2018a}

\section{Concluding Remarks}

These lecture notes have sought to explicate, in a simple way, the principle concepts of dynamical phase transitions. Several archetypal one dimensional topological insulators have been used as examples. The topologically protected edge states have a profound influence on boundary contributions to the return rate which is used to characterise dynamical phase transitions, which can be related to the appearance of special zero eigenvalues in the Loschmidt matrix ${\bm M}(t)$. There also appears to be a little understood relation between entanglement entropy and the critical timescales of dynamical phase transitions. Finally we reviewed some ways in which the Loschmidt amplitude can be generalised to finite temperatures, mixed states, and open systems.

\acknowledgments

This work was supported by the National Science Center in Poland as research Project No.~DEC-2017/27/B/ST3/02881.


%

\end{document}